\documentclass[review,5p,times,twocolumn]{elsarticle}

\usepackage{lineno}

 \usepackage{graphics}

\usepackage{amssymb,amsmath}
\usepackage{array}
\usepackage{multirow}
\usepackage{multicol}
\usepackage{tabularx}
\usepackage{colortbl}
\usepackage{nicefrac}
\usepackage{mathrsfs}
\usepackage{verbatim}
\usepackage[T1]{fontenc}

\usepackage[squaren, binary]{SIunits}

\newcolumntype{Z}{>{\centering\arraybackslash}X}

\biboptions{compress}

\DeclareTextAccent{\myacc}{T1}{4}

\usepackage{lineno}

\biboptions{compress}

\journal{Nuclear Instruments and Methods in Physics Research Section A}

\begin{document}

\begin{frontmatter}

\title{Characterization and performances of DOSION, a dosimetry equipment dedicated to radiobiology experiments taking place at GANIL}

\author[1]{Guillaume Boissonnat}
\author[1]{Jean-Marc Fontbonne\corref{cor1}}
\author[2]{Emmanuel Balanzat}
\author[1]{Frederic Boumard}
\author[1]{Benjamin Carniol}
\author[2]{Amine Cassimi}
\author[1]{Jean Colin}
\author[1]{Daniel Cussol}
\author[1]{David Etasse}
\author[1]{Cathy Fontbonne}
\author[3]{Anne-Marie Frelin}
\author[1]{Jean Hommet}
\author[1]{Jerôme Peronnel}
\author[1]{Samuel Salvador}

\address[1]{LPC (Normandie Univ-ENSICAEN-UNICAEN-CNRS/IN2P3), 6 Bd Mar\'echal Juin, 14050 Caen, France}
\address[2]{CIMAP (CEA/DSM-CNRS/INP-ENSICAEN-UNICAEN), Bd Henri Becquerel, 14076 Caen, France}
\address[3]{GANIL (CEA/DSM-CNRS/IN2P3), Bd Henri Becquerel, 14076 Caen, France}
\cortext[cor1]{Corresponding author - email address: fontbonne@lpccaen.in2p3.fr }

\begin{abstract}
Currently, radiobiology experiments using heavy ions at GANIL (Grand Acc\'el\'erateur National d'Ions Lourds) are conducted under the supervision of the CIMAP (Center for research on Ions, MAterials and Photonics). In this context, a new beam monitoring equipment named DOSION has been developed. It allows to perform measurements of accurate fluence and dose maps in near real time for each biological sample irradiated. In this paper, we present the detection system, its design, performances, calibration protocol and measurements performed during radiobiology experiments. This setup is currently available for any radiobiology experiments if one wishes to correlate one's own sample analysis to state of the art dosimetric references. 

\end{abstract}
\end{frontmatter}
 
\section{Introduction}
\label{sec_intro}

The main advantage of heavy ion therapy compared to conventional radiotherapy or proton therapy lies within its high relative biological effectiveness (RBE) \cite{RBE}. Measuring exact RBE values is still a hot topic in radiobiology as it depends on the particle nature, its linear energy transfer (LET) as well as cellular type. As a nuclear physics research facility, GANIL (Caen, France) produces a wide range of stable nuclei beams starting with carbon ions up to uranium. Nevertheless, delivered beams are not of medical quality and their size are only few millimeters wide. As a result, standard irradiation cell culture flask being 5~$\times$~5~cm$^2$ large, a swept beam must be used to fully irradiate the sample area. In addition, the fluence needed to obtain dose rate of medical interest is several orders of magnitude lower than those of standard physics experiments. Consequently, the low intensity beams used for radiobiology are usually bellow the threshold of standard beam monitors used at GANIL and therefore can vary during the experiment. To overcome these issues, the CIMAP laboratory, in charge of the D1 irradiation room devoted to radiation damage studies and radiobiology, developed a dose calibration protocol for standard radiobiology experiments. It is based on the use of dosimetric films, CR-39 plates and an X-ray monitoring system. Films enable to check the dose delivery homogeneity while CR-39 plates measure the ion fluence (by counting the number of impacts left by ions) and are used to calibrate the X-ray monitor \cite{Durantel}. This dosimetry protocol is robust but time consuming in comparison to experiment's durations (10$\%$). Besides, it does not give access to dose distributions actually delivered for each biological sample. 

Therefore, the LPC developed a few years ago a beam monitor based on an Ionization Chamber (IC) named DOSION for the CIMAP. The first version of DOSION \cite{Dosion1} was a success from a nuclear instrumentation point of view but it was much too complex for day-to-day use. The development in 2008 of a beam monitor dedicated to Pencil Beam Scanning proton therapy for the Belgium company IBA \cite{IC23}, led to the improvement of DOSION. This new DOSION aims to simplify and shorten the dose calibration for experiments at GANIL and to produce an accurate dose map for each irradiated biological sample. In this work, we present this new version of DOSION and its calibration procedure as performed during the 2014 BioGraphic experiment (part of the France Hadron~\cite{FrHadron} French national infrastructure on particle therapy research). BioGraphic is a joint collaboration between radiobiologist teams from the CYCERON imaging and biomedical research facility and from the François Baclesse cancer treatment center as well as physicists from LPC. The goal is to irradiating cancerous and normal cells of different types in the same conditions and at different LET (from 28 to 75~keV$\cdot\mu$m$^{-1}$) to measure and compare biological effects (RBE, DNA lesions, survival rates). Both radiobiology teams are studying cancerous cells and their healthy tissue counterparts, respectively glyoblastoma at CYCERON \cite{Peres} and fibroblastoma \cite{Laurent} at Baclesse. 

\section{Material and methods}
\subsection{Cell irradiation protocol}

In order to optimize the irradiation of biological samples, the D1 room at GANIL is equipped with an automatic sampler that can hold up to twenty four 5~$\times$~5~cm$^2$ cell culture flasks. Each individual sample typically receives between 0.5 to 8~Gy, at a standard 2~Gy$\cdot$min$^{-1}$ dose rate, resulting in the individual irradiation time to last between 30~s to 4~min. Overall, a full irradiation sequence of 24 samples can last up to an hour, taking in consideration the room entrance and exit protocol durations.

During each sample irradiation, CIMAP's X-ray counter monitors the number of ions delivered and stops the beam when the specified fluence is met before starting it again when the next sample has been placed in the irradiation position. Simultaneously, DOSION aims to provide an accurate dose map of each irradiation in the sequence, therefore adding spatialization to CIMAP's ions count. 

\subsection{Beam conditions and experimental setup}

\begin{figure}[!ht]
\begin{center}
\includegraphics[angle=0, width=\columnwidth]{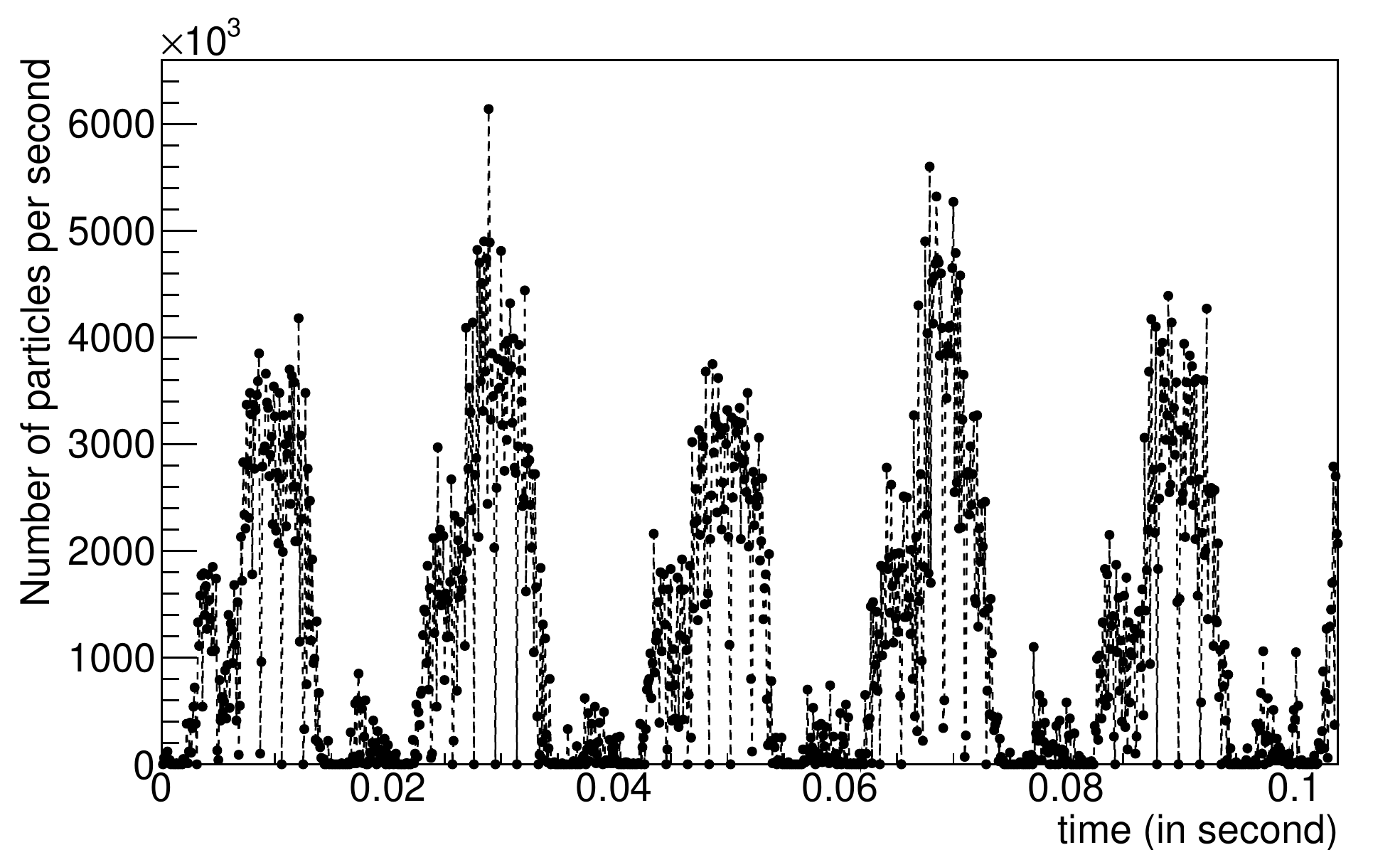}
\end{center}
\caption{Beam intensity for a fixed beam as a function of time (sampled at 10~kHz) with an averaged intensity of 1.2$~\times$~10$^6$ $^{12}$C$\cdot$s$^{-1}$.}
\label{intensity_t}
\end{figure}

The Gaussian shaped beams delivered by GANIL was moved across the 5~$\times$~5~cm$^2$ irradiation field by two sets of sweeping magnets, at 400~Hz and 4~Hz on $X$ and $Y$ axis, respectively. In addition, the irradiation field rims were refined by the use of tungsten jaws. 
While the standard beam intensity used at GANIL for physics experiments is of about $10^{10}$~particles$\cdot$s$^{-1}$, the one needed for radiobiology experiments was much lower: between $5\times10^6$ and $5\times10^7$~particles$\cdot$s$^{-1}$. This low beam intensity was obtained using slits ("pepperpots") to cut portions of the nominal intensity beam. This intensity reduction process had the unfortunate consequence to display a somewhat chaotic instantaneous intensity (see Figure~\ref{intensity_t}). In fact, little fluctuations of the beam position in the slits plane can cause the instantaneous intensity to change by a few orders of magnitude. 
 
Those fast fluctuations can sometimes lead to large inhomogeneities in the irradiation field within the few seconds irradiations~\cite{Dosion1}. This effect is suppressed by averaging it across the irradiation field, the beam was widen from 1.5~mm to 3~mm (in standard deviation) by a 2~$\mu$m titanium foil placed nine meters before the target.

During the BioGraphic experiment a native $^{12}$C beam at 95~MeV$\cdot$A$^{-1}$ was used. Several Polymethylmethacrylate (PMMA) energy degrader were inserted in the beam to study multiple energies and therefore the impact of their respective LET on the RBE in a single experiment by saving the beam energy tuning time. PMMA degraders thicknesses are presented in Table~\ref{table:BeamConf} as well as calculated resulting energy and LET at which the beam met the cells. 

\begin{table}[!ht]
\caption{Beam configurations using native beams of $^{12}$C at 95~MeV$\cdot$A$^{-1}$ and corresponding calculated beam energies and LET in the cells.}
\centering
\footnotesize
\label{table:BeamConf}
\renewcommand{\arraystretch}{1.3}
{
	\begin{tabularx}{\linewidth}{ZZZZ}
	\hline
	\hline
	PMMA thickness (mm) & Calculated energy in the cells (MeV$\cdot$A$^{-1}$)	& Calculated LET in the cells (keV$\cdot\mu$m$^{-1}$)\\
	\hline
	0	&	91.8 & 28.2\\
	6.9 	&	71.1	& 34.2\\
	13.9	&	43.5	& 50.5\\
	16.9 &	25.7	& 77.3\\
	\hline
	\hline
	\end{tabularx}
}
\end{table}

\begin{figure*}[!t]
\begin{center}
\includegraphics[angle=0, width=\textwidth]{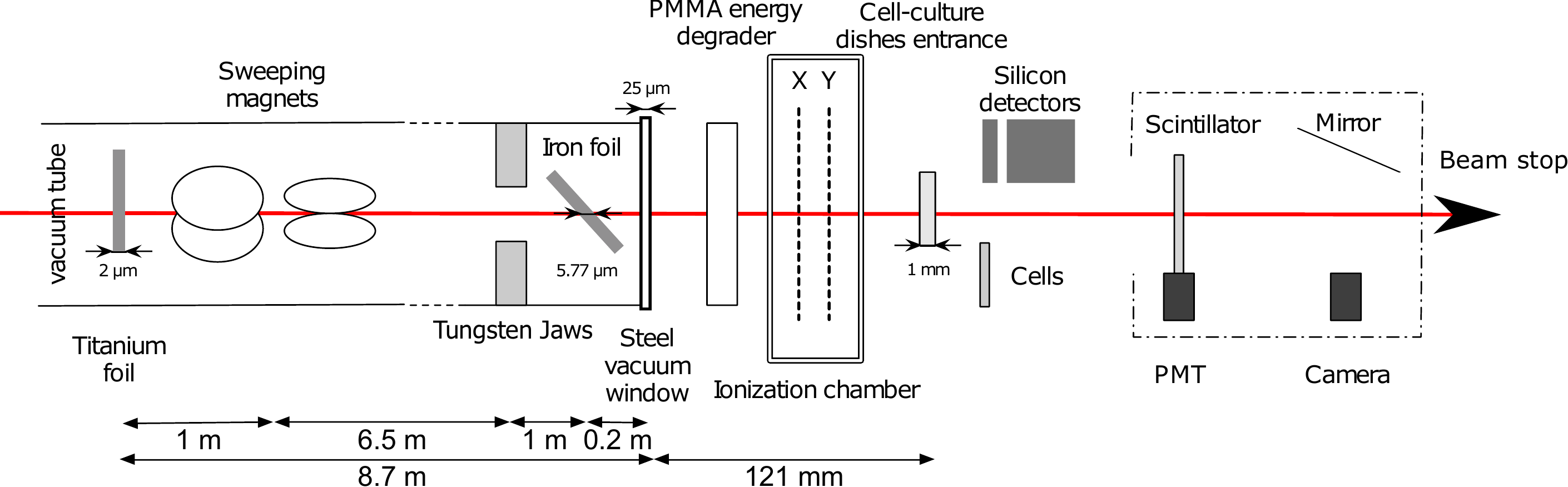}
\end{center}
\caption{Bean line and DOSION calibration set up.}
\label{BeamLineSchem}
\end{figure*}

The overall irradiation setup is shown in Figure~\ref{BeamLineSchem}. It was composed of a 2~$\mu$m titanium foil to increase the beam size, a 5.77~$\mu$m iron foil for the X-ray monitoring system, a 25~$\mu$m steel plate to separate the beam line vacuum pressure from the irradiation room, a PMMA energy degrader, the DOSION ionization chamber and finally a biological sample put in a polystyrene cell culture flask with a 1~mm entrance window. During the calibration procedure, two thin silicon detectors replaced the biological sample for LET measurements and a plastic scintillator coupled to a photomultiplier tube (PMT) was placed further away to calibrate the ionization chamber in number of ions. In addition, a CCD camera was used to visualize the beam shape and irradiation field on the scintillator using a mirror.

\subsection {DOSION ionization chamber}

\begin{figure}[!ht]
\begin{center}
\includegraphics[angle=0, width=0.9\columnwidth]{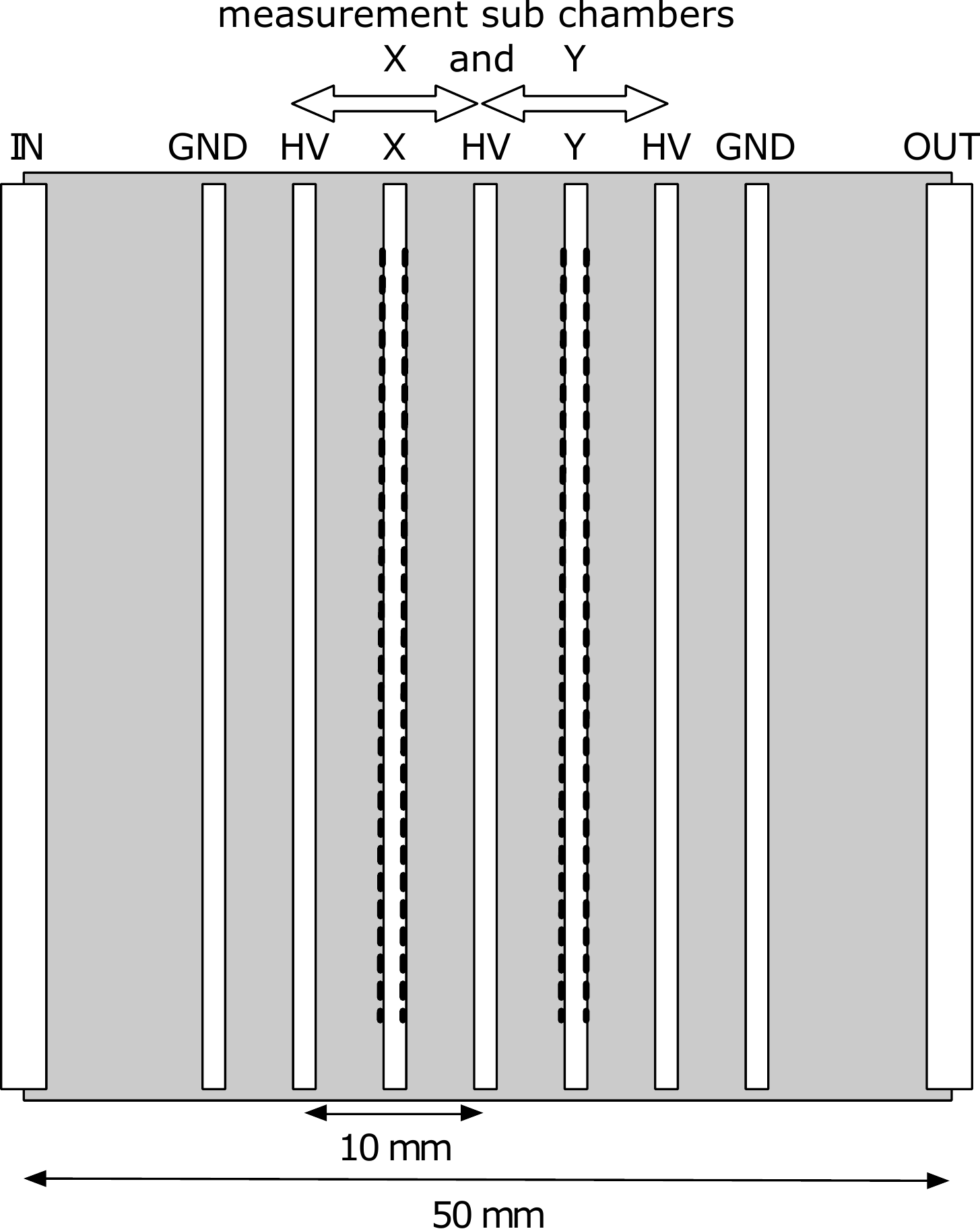}
\end{center}
\caption{Layout of the ionization chamber DOSION.}
\label{DosionLayoutA}
\end{figure}

\begin{figure}[!ht]
\begin{center}
\includegraphics[angle=0, width=0.9\columnwidth]{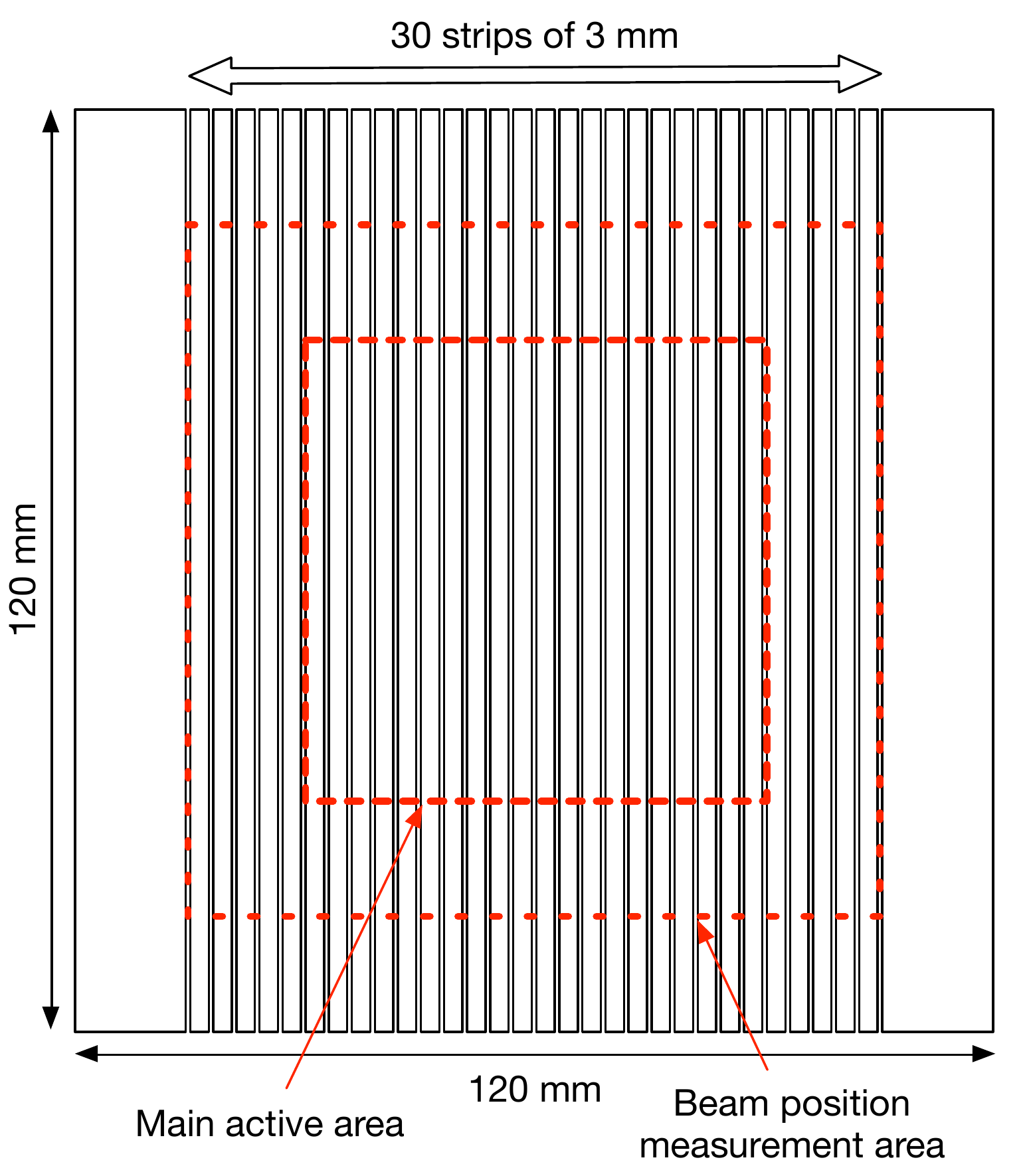}
\end{center}
\caption{Layout of the stripped electrodes.}
\label{DosionLayoutB}
\end{figure}

This upgraded version of DOSION was based on the development of IBA's dedicated PBS ionization chambers previously made at LPC for IBA's \cite{IC23}.  These 30~$\times$~30~cm$^2$ parallel plate air ionization chambers consisted in two sub-chambers measuring the beam intensity and two stripped sub-chambers measuring the position and the size of the beam along x and y axes.
In our cases, considering that medical precision and redundancy were not needed, the design was adapted to get a 6~$\times$~6~cm$^2$ main active area with only two stripped sub-chambers with two 5-mm air gaps (see Figure~\ref{DosionLayoutA}). Each measurement electrode was divided into 32 strips of length 120~mm: 2 large strips (15~mm wide) on the sides and 30 strips of 3~mm in the center (see Figure~\ref{DosionLayoutB}). The summed signal of the 32 strips was used for dose measurements and the 30 central strips individual signals were used for beam localization and shape measurement in a 9~$\times$~9~cm$^2$ area. This new design made the chamber smaller while offering a theoretical spatial resolution better than 10~$\mu$m for beams larger than 1.5~mm.

Stripped measurement electrodes were made of 2.5~$\mu$m Mylar foils with 170~nm gold coating on each side. All other electrodes are all made of Mylar with 200~nm aluminum coating, the respective Mylar thickness are 1.5~$\mu$m for both high voltage and ground electrodes and 12~$\mu$m for both entrance and exit windows.
Signal measurement were conducted using the in-house build acquisition system FASTER~\cite{faster} with a two CARAMEL daughter cards setup (based on the Texas Instrument DDC316 electrometer chip~\cite{DDC316}) for a total of 64 electrometer channels. The system ensured measuring simultaneously the 64 strips of DOSION at a maximal rate of 25~kHz (i.e. 1.6 million samples per second).
Each channel could measure up to 24~pC per sample, which was much more than needed considering that a standard 2~Gy$\cdot$min$^{-1}$ irradiation rate would lead to a 1~pC signal per 40~$\mu$s sample time. 

\begin{figure}[!ht]
\begin{center}
(a)\includegraphics[angle=0, width=0.9\columnwidth]{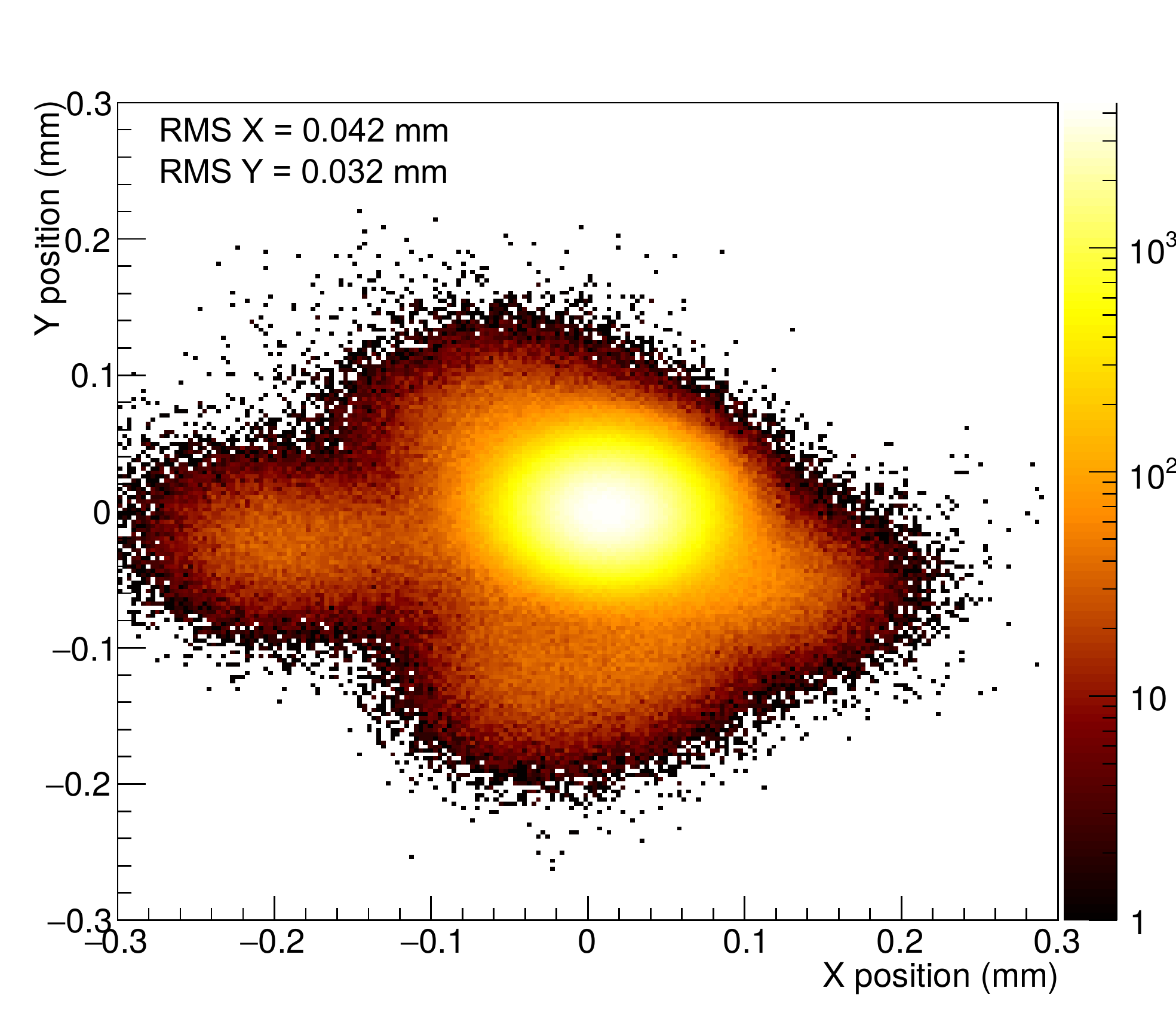}
(b)\includegraphics[angle=0, width=0.9\columnwidth]{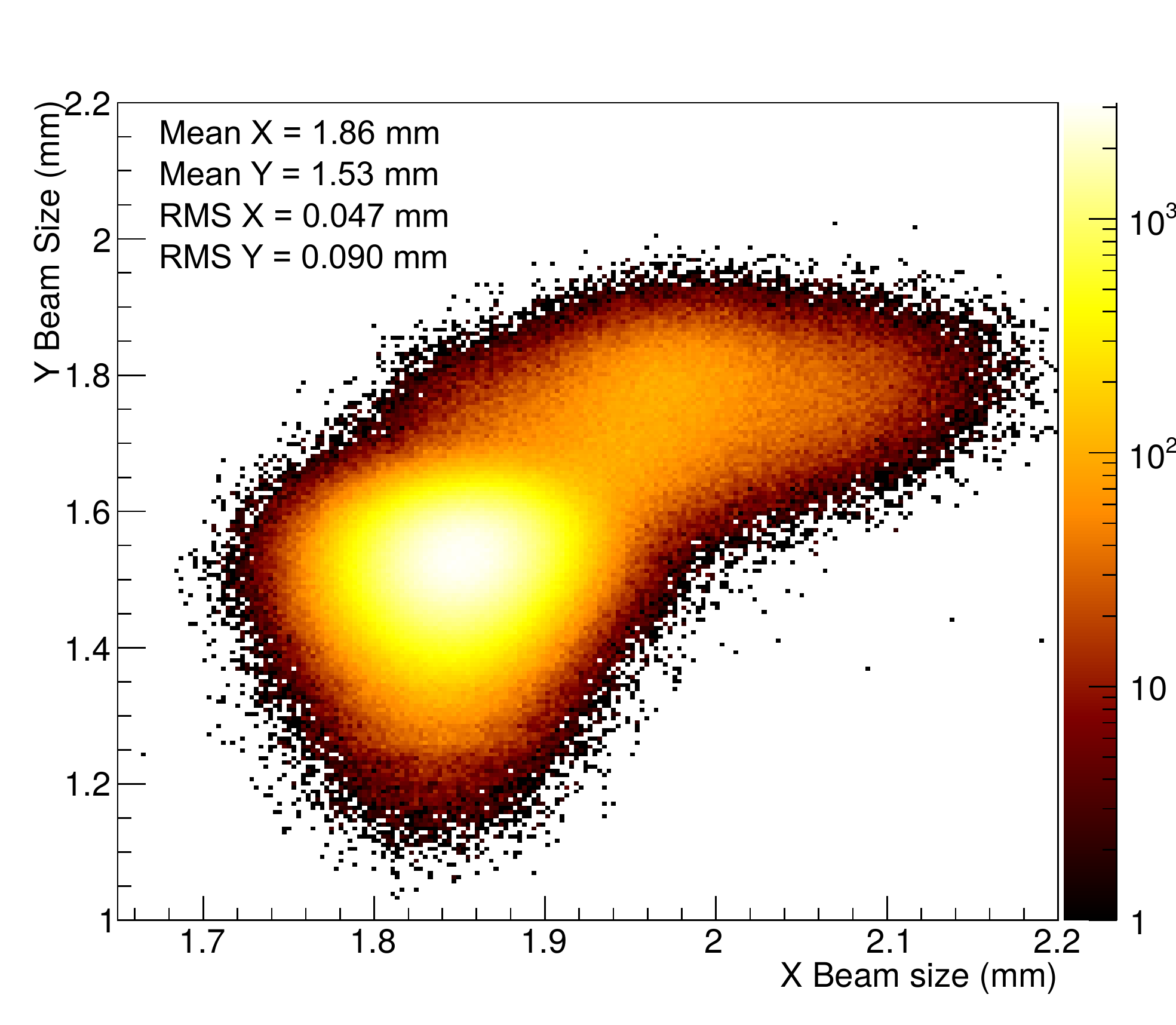}
\end{center}
\caption{Beam position (a) and size measured (b) during 100~s with a 25~kHz sampling rate at ARRONAX using a fixed beam of 17~MeV$\cdot$A$^{-1}$ alpha particles at a 20~pA intensity.}
\label{Arronax}
\end{figure}

The raw performances in terms of spatial resolution were obtained at the ARRONAX facility (Saint-Herblain, France) using a C-70 IBA accelerator~\cite{C70} which was more suited than GANIL for such tests. Those measurements were performed using a fixed beam of 17~MeV$\cdot$A$^{-1}$ alpha particles. Spatial resolution performances were obtained by fitting the position and the size of the beam at a 25~kHz rate with an intensity of 20~pA. As presented in Figure~\ref{Arronax}, beam position and size distributions made at the ARRONAX facility have uncertainties of about 40~$\mu$m$_{RMS}$ on both position and beam size. These measurements were consistent with the expected 10~$\mu$m resolution while the actual beam stability is unknown.

In order to measure the response uniformity of the chamber, a Mini-X X-ray generator~\cite{MiniX} was used, monitored by an X123-CdTe X-ray spectrometer~\cite{X123}.The X-ray generator was used at 50~kV and 79~$\mu$A, hardened with a 1~mm aluminum filter and collimated with a 2~mm radius brass collimator to get a 3~mm wide Gaussian shaped beam on the ionization chamber. The uniformity was measured as the ratio between the X-ray spectrometer count and the ionization chamber signal.  The X-ray source and the CdTe were aligned in a fixed position while the chamber was moved to change the position of irradiation. The 6~$\times$~6~cm$^2$ main active area and the 9~$\times$~9~cm$^2$ overall localization field of the ionization chamber were respectively scanned with 15~mm and 22.5~mm steps, resulting in respecting homogeneities for both fields of 0.8$\%_{RMS}$ (3$\%_{peak-to-peak}$) and 1.3$\%_{RMS}$.

\subsection {Sensitivity measurement of the IC}

Before being able to obtain a dose measurement, the IC sensitivity must be calibrated against the number of incident ions. Despite the fact that the ion sensitivity of the chamber could be estimated using Eq.~\ref{eqSensitivity}, the uncertainties on the LET inside the ionization chamber and after the PMMA (and in a lesser way on the gap size $d$, the air density $\rho_{air}$ and the ionization potential $W_{air}$~\cite{ICRU}) made an absolute calibration mandatory to get calibration coefficients with a precision better than few percents.

\begin{equation}
{sensitivity}_{IC}= \frac{LET_{IC} \cdot d}{W_{air} \cdot \rho_{air} }
\label{eqSensitivity}
\end{equation}

\begin{figure}[!ht]
\begin{center}
\includegraphics[angle=0, width=\columnwidth]{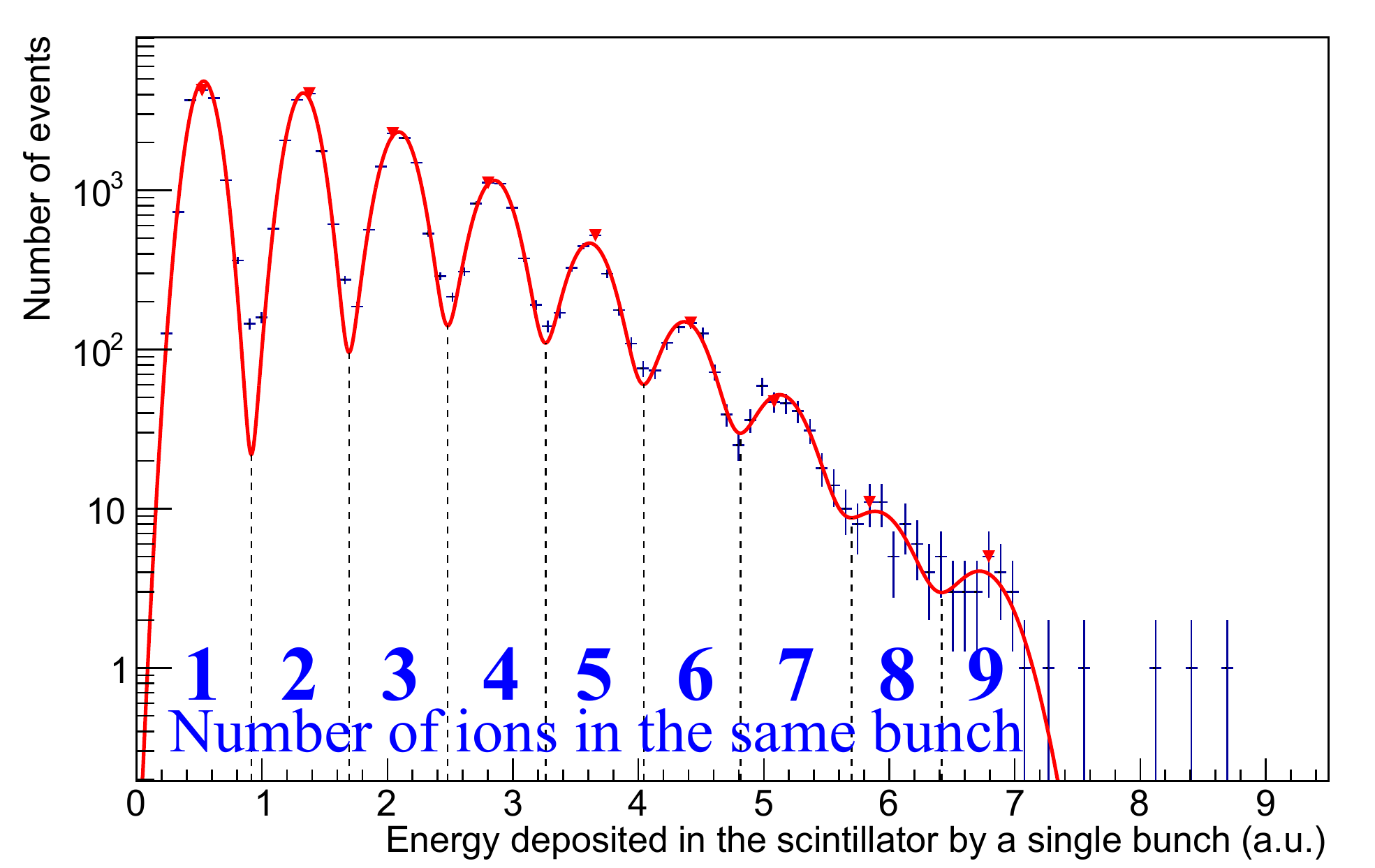}
\end{center}
\caption{Distribution of the energy deposited in the scintillator by an ion bunch and energy thresholds used for ion count.}
\label{IonPerBunch}
\end{figure}

Absolute calibration was then performed using a plastic scintillator coupled to a PMT placed after the chamber in the beam line to count the number of ions impinging the chamber. GANIL produces only mono-energetic particle beams which would result in the same amount of energy deposited in the scintillator, However, ions are delivered in few nanoseconds bunches causing the number of ions per bunch to follow a near Poisson distribution. In our case, the number of detected ions is shown in Figure~\ref{IonPerBunch}, each Gaussian shaped peak corresponds to a certain number (from one to nine) of ions in the same bunch or event (when looking at it from the detector point of view). Thresholds were then specified by the user for counting the number of ions per bunch, represented as dashed lines.
In addition, the ion sensitivity measurements was conducted using fixed beams to limit the light collection dependency on the position between the scintillator and the PMT. The signal was acquired using a FASTER charge integration scaler daughter board enabling synchronized measurement with the ionization chamber.

\subsection {LET measurements}

In order to obtain a reliable dose measurement, the linear energy transfer of the particles in the cells had to be measured. The LET measurements were conducted using two thin silicon detectors assembled as a $\Delta$E-$\Delta$E telescope with thicknesses of 148~$\mu$m and 1.04~mm, respectively. A standard three-alpha source $^{239}$Pu,$^{241}$Am and $^{244}$Cm was used to calibrate both detectors.
The beam energy reaching the biological cells is deduced from the energy loss in the silicon detectors knowing their thicknesses. The LET was also calculated for that beam energy using Bethe-Block formula and Geant4 simulations.
Preamplified output signals were read out and digitalized using a FASTER daughter board dedicated to spectrometry (named MOSAHR).

\subsection {Fluence map and dose distribution}

When the energy deposited by each incoming particle was known (through the $LET$ in keV$\cdot \mu$m$^{-1}$), the delivered dose $D$ (in Gy) can be calculated using Eq.~\ref{eqDose}, where $\Phi$ is the beam fluence (in number of ions per mm$^{2}$) and $\rho_{water}$, the density of the cell culture medium, approximated as water (in g$\cdot$cm$^{-3}$). 

In the previous version of DOSION, the fluence map was created using the beam intensity and position at every step to create a first map and then convolving the resulting image with the Gaussian beam shape. The beam intensity was measured using an ionization chamber, the beam position using the sweeping magnetic field measured by Hall effect sensors and finally the beam shape was measured using its image on a scintillator through the CCD camera.

\begin{equation}
D = 1.602 \cdot 10^{-7}~\frac{LET\cdot \Phi }{\rho_{water} }
\label{eqDose}
\end{equation}

The complexity of that process made it very little user-friendly. Each parameter can now be measured simultaneously with the ionization chamber. The production of a fluence map was done by measuring the position, the beam shape and the intensity at every time step (40~$\mu$s) and adding the corresponding fluence distribution to the summed fluence map. However, due to a low beam intensity, its fluctuations made it difficult to differentiate the beam from the acoustic noise. If it was in theory possible to perfectly track the beam knowing the sweeping magnets frequencies, a simpler and safer approach was implemented. The sampling was increased from 40~$\mu$s to 2.4~ms corresponding to a single sweep on $X$-axis and to a 1~mm displacement on the $Y$-axis. With such measurements, a simple back projection of $X$ and $Y$ profiles created a fluence map while keeping the relevant spatial information on the delivered dose. Nevertheless, it had the drawback to limit by construction the spatial resolution to the 3~mm strip size. The delivered dose distribution was extracted from the fluence map, enabling the radiobiology teams to decide whether it was homogeneous enough to be considered in their studies.

The current process of dose reconstruction was fairly simple: if the number of ions seen in a 2.4~ms sample is higher than 100 (about 0.1~pC depending on the beam energy), the $X$ and $Y$ profiles registered during that sample time were back projected and added to the total fluence map. Otherwise, the profiles are added to the "lower than threshold" $X$ and $Y$ profiles. At the end of each irradiation those last profiles were also back projected and added to the summed fluence map so that no dose was lost in the process while minimizing the impact of noise. Once the fluence map was created, a 5~$\times$~5~cm$^2$ area was delimited and the inside dose distribution was plotted. In this dose distribution, a 3~mm on each side of the box were removed so that the sharpness of the irradiation field (due to the use of tungsten jaws) was not mistaken for dose inhomogeneities.

\section{Results}

\subsection {Sensitivity measurement of the IC}

Measurements and calculations of the sensitivity are presented in Table~\ref{table:Sensitivity}. Regardless the good agreement between the measured and calculated sensitivity, we observed a deviation of about 3$\%$ which might be caused by an underestimation of the gap size (about 150~$\mu$m out of the 5~mm air gap).

\begin{table}[!ht]
\caption{Measurements and calculations (Bethe-Block) of DOSION IC's sensitivity to $^{12}$C for different PMMA energy degraders.}
\centering
\footnotesize
\label{table:Sensitivity}
\renewcommand{\arraystretch}{1.3}
{
	\begin{tabularx}{\linewidth}{ZZZZ}
	\hline
	\hline
	PMMA thickness & Measured sensitivity & Calculated sensitivity & Deviation\\
	\hline
	0~mm	&	1.36~fC/$^{12}$C		&	1.32~fC/$^{12}$C	 	& +3.0$\%$ \\
	6.9~mm	&	1.63~fC/$^{12}$C		&	1.58~fC/$^{12}$C 		& +3.2$\%$ \\
	13.9~mm	&	2.31~fC/$^{12}$C		&	2.22~fC/$^{12}$C		& +4.1$\%$ \\
	16.9~mm	&	3.15~fC/$^{12}$C		&	3.05~fC/$^{12}$C		& +3.3$\%$ \\
	\hline
	\hline
	\end{tabularx}
}
\end{table}

\subsection {LET measurements}

The $^{12}$C linear energy transfer in water measured from the energy lost in the two silicon detectors is presented along with Bethe-Block calculations and Geant4 simulations in Table~\ref{table:LET}. Calculations are in good agreement with the measurements. However, the deviation increases with the degrader thickness. This is mostly due to the uncertainty on the PMMA thicknesses and density as a small error on these values would lead to a large error on the results. This error increases when the beam energy approaches Bragg peak.

\begin{table}[!ht]
\caption{LET measured in the cell culture flask for $^{12}$C using different PMMA energy degraders along with calculations using Bethe-Block equation and Geant4 simulations.}
\centering
\footnotesize
\label{table:LET}
\renewcommand{\arraystretch}{1.3}
{
	\begin{tabularx}{\linewidth}{ZZZZ}
	\hline
	\hline
	PMMA thickness & Measured LET (keV$\cdot\mu$m$^{-1}$) & Calculated LET (Bethe-Block) & Simulated LET (Geant4) \\
	\hline
	0~mm 	&	28.2~$\pm$~0.5 	& 28.2~keV$\cdot\mu$m$^{-1}$ & 28.3~keV$\cdot\mu$m$^{-1}$\\
	13.9~mm	&	49.8~$\pm$~0.6	& 50.5~keV$\cdot\mu$m$^{-1}$ & 48.1~keV$\cdot\mu$m$^{-1}$\\
	16.9~mm	&	73.4~$\pm$~1.1	& 77.3~keV$\cdot\mu$m$^{-1}$ & 70.8~keV$\cdot\mu$m$^{-1}$\\
	\hline
	\hline
	\end{tabularx}
}
\end{table}

The silicon detectors were also used to qualitatively investigate the pollution due to fragmentation processes of $^{12}$C in the PMMA energy degraders or other elements in the beam line, as well as the cell culture flask entrance window. The estimated fragmentation pollution is presented in Table~\ref{table:Pollution} along with the estimated LET of the two most frequent fragments, $Z=1$ and $Z=2$ particle(s). In addition, the dose deviation due to fragmentation has been roughly estimated considering that all fragments had alpha particles LET. The dose deviation calculated as such being lower than 1$\%$, the impact of fragmentation processes is neglected in the following.

\begin{table}[!ht]
\caption{Secondary particles pollution estimated in the cell culture flask for $^{12}$C using different PMMA energy degraders.}
\centering
\footnotesize
\label{table:Pollution}
\renewcommand{\arraystretch}{1.3}
{
	\begin{tabularx}{\linewidth}{ZZZZZ}
	\hline
	\hline
	PMMA thickness & Fragments fraction & Proton LET (keV$\cdot\mu$m$^{-1}$)	& Alpha LET (keV$\cdot\mu$m$^{-1}$)		& Dose deviation \\
	\hline
	0~mm	&	0.5$\%$		& 					&				& 					\\
	13.9~mm	&	5.5$\%$		& 1					& 4		&	0.4$\%$			\\
	16.9~mm	&	6.5$\%$		& 1			 		& 5		&      0.5$\%$			\\
	\hline
	\hline
	\end{tabularx}
}
\end{table}

\subsection {Fluence map and dose distribution}

\begin{figure}[!h]
\begin{center}
(a)\includegraphics[angle=0, width=\columnwidth]{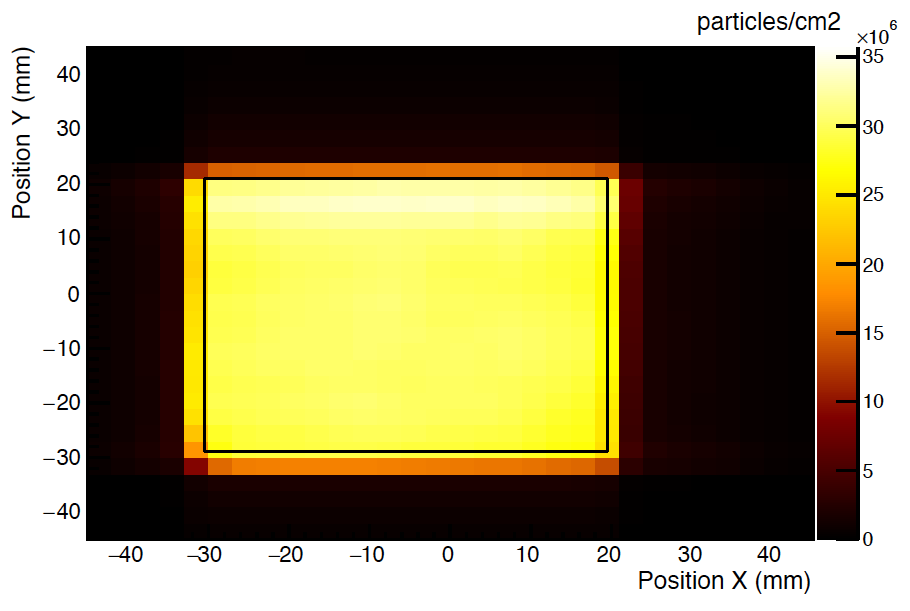}
(b)\includegraphics[angle=0, width=0.95\columnwidth]{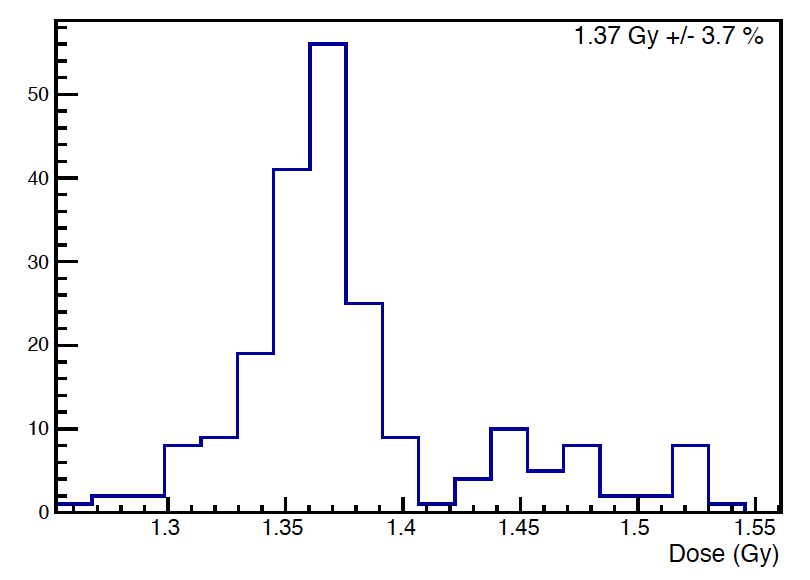}
\end{center}
\caption{Fluence map (a) and dose distribution (b) in the cell culture flask (in number of pixel at each dose) obtained for a one minute long 95~MeV$\cdot$A$^{-1}$ $^{12}$C irradiation at GANIL.}
\label{fluenceMap}
\end{figure}

During the 2014 BioGraphic experiments, as presented in Table~\ref{table:ResSum}, most of the irradiations had homogeneities better than 4$\%_{RMS}$ and the dose dispersion between samples lower than 2$\%_{RMS}$ for the two lower LET and 4$\%_{RMS}$ for the highest one. An example of such irradiation is presented in Figure~\ref{fluenceMap}. 

\begin{table}[!ht]
\caption{Summarized results of the 2014 BioGraphic experiment, presenting for every energy degraders the averaged fluence normalized by the delivered dose, the dose dispersion between samples and the averaged spatial homogeneity.}
\centering
\footnotesize
\label{table:ResSum}
\renewcommand{\arraystretch}{1.3}
{
	\begin{tabularx}{\linewidth}{ZZZZZZ}
	\hline
	\hline
	PMMA thickness & Fluence/dose (cm$^{-2}\cdot$Gy$^{-1}$)	& Dose dispersion 	& Spatial homogeneity & N$_{samples}$	\\
	\hline
	0~mm	&				2.2$\times$10$^7$		& 	 	1.4$\%$		&  3.8$\%$ & 	84	\\
	13.9~mm	&				1.3$\times$10$^7$		& 	 	1.4$\%$		&  4.0$\%$ &	72	\\
	16.9~mm	&				8.4$\times$10$^6$		& 		3.9$\%$		&  4.5$\%$ &	78	\\
	\hline
	\hline
	\end{tabularx}
}
\end{table}

\section{Discussion}

Compared to the previous version of DOSION, the spatial resolution of the fluence maps has been degraded as the back projection algorithm limits the resolution to the strip size (3~mm). Many algorithms were tested to track the beam and enhance the fluence map reconstruction to get a spatial resolution closer to the 40~$\mu$m measured on fixed beam. The first issue was the beam position tracking (using the 25~kHz sampling). While the intensity was fluctuating a lot, it was likely to have beam-on samples bellow the noise threshold. If a single sample can be neglected in regard to the overall dose, the sum of them cannot. Therefore, a reliable beam position tracker had to be implemented to infer the beam position even for low intensity beams. If the sweeping frequency was supposed to be very stable at respectively 4 and 400~Hz, those frequencies were not stable enough to be used as such and had to be carefully tracked. In addition, the beam position suffered from 50~Hz pollution as it was observed on supposedly fixed beams with peak-to-peak displacement of few millimeters that render safe position tracking nearly impossible. The current use of lead jaws had also the undesired effect to change drastically the beam shape, not only by cutting its edges but also by changing the beam shape through diffusion. Therefore the exact beam shape had to be calculated for every position of the beam even if the central beam positions were located behind the lead jaws.
All of this made a precise fluence map reconstruction possible but at the same time very dependent on the irradiation conditions and therefore subject to reliability issues. Such approach could be implemented if DOSION was to be used in facilities with more stable beams and reliable beam positioning.
In fact, the new DOSION ionization chamber has the advantage to be much more versatile than its predecessor as it has already been tested at ARRONAX facility and at the ORSAY Proton therapy center with satisfying results. 

As a whole, the DOSION ionization chamber along with its calibration process should allow to get precise measurements of fluence and dose delivery for every cell irradiation at GANIL. In addition, if calibration is not possible, default calibration can be inferred from previous experiments or simple calculations with uncertainties better than 5$\%$. 

\section{Conclusion and perspectives}

In this paper, we have demonstrated that it was possible to use beam energy degraders while keeping good knowledge on LET and on the dose delivered while sparing the dozen of hours needed at GANIL to change the beam energy. In addition, the calibration time of DOSION is fully compatible with CIMAP dosimetric procedure and therefore should be absolutely transparent from the user point of view. 
During the year 2014, the use of DOSION highlighted a focusing magnet failure that had catastrophic consequences on the dose distribution and was undetectable by CIMAP's X-ray counter. 
The new DOSION version has been and is used on hundreds of cell cultures irradiations and seems to respond well to radiobiologist needs.

\section{Acknowledgements}

The authors wish to thank the Basse Normandie Region for funding G. Boissonnat's PhD thesis on the behalf of ARCHADE, France Hadron for dedicating beam time at GANIL to test DOSION prior to radiobiology experiment; CIMAP's teams for giving the opportunities to participate to their dose calibration and lending us time to performed our own calibration of DOSION; radiobiologist teams from Baclesse, CYCERON and LARIA~\cite{Hamdi} for allowing DOSION's validation during their experiments; ARRONAX and CPO teams for enabling us to evaluate the ionization chamber performances on their installations; LPC's instrumentation and mechanical teams for building the ionization chamber as well as adapting it to GANIL D1 beam line and LPC's FASTER and electronic teams for developing the CARAMEL daughter card and DOSION's user interface.

\bibliography{MyBib}

\end{document}